\documentclass[twoside,a4paper,11pt]{sea22}
\usepackage{graphicx}
\usepackage{hyperref}
\usepackage{media9}
\begin{document}
\pagenumbering{arabic}
\pagestyle{myheadings}
\thispagestyle{empty}
{\flushleft\includegraphics[width=\textwidth,bb=58 650 590 680]{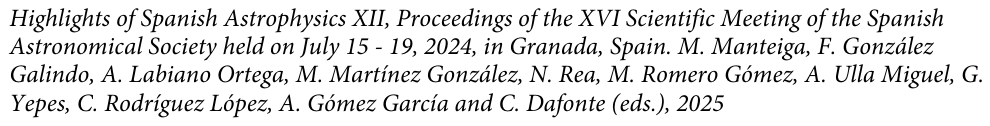}}
\vspace*{0.2cm}
\begin{flushleft}
{\bf {\LARGE
%
GPU-Accelerated Searches for Long-Transient Gravitational Waves from Newborn Neutron Stars
%
}\\
\vspace*{1cm}
%
Mérou, J. R.$^{1}$,
Tenorio, R.$^{1,\,2,\,3}$
and Sintes, A. M.$^{1}$
%
}\\
\vspace*{0.5cm}
%
$^{1}$
Departament de Física, Universitat de les Illes Balears, IAC3 - IEEC, Carretera de Valldemossa km 7.5, E-07122 Palma, Spain\\
$^{2}$
Dipartimento di Fisica “G. Occhialini”, Universitá degli Studi di Milano-Bicocca, Piazza della Scienza 3, 20126 Milano, Italy\\
$^{3}$
INFN, Sezione di Milano-Bicocca, Piazza della Scienza 3, 20126 Milano, Italy
%
\end{flushleft}
%
\markboth{
GPU-Accelerated Searches for tCWs from Newborn NSs
}{ 
%
J. R. Mérou et al.
%
}
\thispagestyle{empty}
\vspace*{0.4cm}
\begin{minipage}[l]{0.09\textwidth}
\ 
\end{minipage}
\begin{minipage}[r]{0.9\textwidth}
\vspace{1cm}
\section*{Abstract}{\small
%
We present a novel method to efficiently search for long-duration gravitational wave transients emitted by new-born neutron star remnants of binary neutron star coalescences or supernovae. The detection of these long-transient gravitational waves would contribute to the understanding of the properties of neutron stars and fundamental physics. Additionally, studying gravitational waves emitted by neutron stars can provide valuable tests of general relativity and offer insights into the neutron star population, of which only a small fraction appears to be observable through current electromagnetic telescopes. Our approach uses GPUs and the JAX library in Python, resulting in significantly faster processing compared to previous methods. The efficiency of this code enables wide regions of the sky to be covered, eliminating the need for precise pinpointing of mergers or supernovae. This method will be deployed in searches for long-transient gravitational waves following any detection of a binary neutron star system merger in the latest O4 science run of the LIGO-Virgo-KAGRA collaboration, which started in May 2023 with a significant improvement in sensitivity with respect to previous runs.
%
\normalsize}
\end{minipage}
%
%
%
\section{Introduction \label{intro}}

Gravitational Waves (GWs) are perturbations on the geometry of space-time produced by accelerated asymmetric distributions of mass or energy \cite{misner73}, first predicted by Albert Einstein in 1916 \cite{einstein16}. The first detection of GWs took place in 2015, by the laser interferometers of the LIGO Scientific Collaboration \cite{abbot16}. Since then, GWs have been detected from the mergers of binary black holes and neutron stars \cite{abbot23}. A notable event was the first ever observed merger of a binary neutron star system, GW170817. This event was a breakthrough in multi-messenger astronomy, since it was also seen through electromagnetic observatories \cite{abbot17}.

Neutron stars (NSs) are dense remnants formed from the core collapse of stellar supernovae in stars with masses between 8 and 30 times that of the sun. Many aspects of them, such as their equation of state, inner structure, and their role in the synthesis of heavy elements remain not fully understood \cite{lattimer2004}. Studying GWs emitted by NSs can provide tests of general relativity and offer insights into the NS population, a majority of which appears to be currently undetectable by existing electromagnetic telescopes \cite{reed21}.

The detection of GW170817 initiated efforts within the GW research community to develop search methods designed to detect GWs originating from the remnants of such mergers \cite{abbot18}. Theoretical predictions suggest that these remnants, potentially NSs with high ellipticity and rapid spin rates, could emit long-duration transient gravitational waves (tCWs). Detecting these tCWs would contribute to our understanding of the properties of NSs and fundamental physics \cite{abbot21}. 

To model the frequency evolution of a tCW, we use the msMagnetarWaveform model \cite{lasky17}, in which the GW frequency is
\begin{equation}
    f_{\textrm{gw}}\left(t\right)=f_{\textrm{gw},0}\left(\frac{t-T_{0}}{\tau}+1\right)^{\frac{1}{1-n}}\ ,
\end{equation}
where $f_{\textrm{gw},0}$ is the initial emission frequency at  time $t=T_{0}$, $\tau$ denotes the spindown timescale, and $n$ is referred to as the braking index, which characterizes the emission mechanism of the GW. A braking index of $n=5$ corresponds to pure GW emission from a non-axisymmetric rotator \cite{alford12}. The theoretical ranges for $f_{\textrm{gw,0}}$, span from 500 Hz to 3000 Hz, 3500 s to 35000 s for $\tau$, and 3 to 7 for $n$ \cite{lasky17}.

The instantaneous signal frequency in the detector frame is given by the GW frequency adjusted for a Doppler modulation,
\begin{equation}
    \hat{f}_{\textrm{gw}}\left(t\right)=f_{\textrm{gw}}\left(t\right)\left(1+\frac{\vec{v}\left(t\right)\cdot\vec{n}}{c}\right)\ ,
\end{equation}
where $\vec{v}$ is the detector's velocity relative to the solar system barycenter, $c$ is speed of light, and $\vec{n}$ the sky location of the source.

The strain amplitude $h_0(t)$ produced by the GW in the detector's output depends on the emission mechanism. For a constant non-axisymmetric deformation of the source NS, it evolves as
\begin{equation}
    h_{0}\left(t\right)=\frac{4\pi^{2}G}{c^{4}}\frac{I_{zz}\epsilon}{d}f_{\textrm{gw}}^{2}\left(t\right)\ ,
\end{equation}
where $G$ is the gravitational constant, $I_{zz}$ the $z-z$ component of the star's moment of inertia, set to  $4.34\times10^{38}$ $\textrm{kg\ m}^{2}$ as in \cite{oliver19}, $d$ the distance between the detector and the source and $\epsilon$ the equatorial ellipticity of the star \cite{zimmerman79}.

\section{Analysis method \label{analysis_method}}

None of the developed search methods for tCWs have yet achieved any detection. One of the developed methods is the Adaptive Transient Hough method (ATrHough) introduced by Oliver et al. \cite{oliver19}. We have developed a new search method that surpasses the capabilities of ATrHough, achieving similar sensitivities with significantly lower computational costs. 

This new method makes use of the power of GPU acceleration, resulting in much faster processing times. The code has been implemented using JAX, a Python-based library that allows for GPU computation and the just-in-time (JIT) compilation for optimized performance \cite{jax2018github}.

\begin{figure}[!h]
    \center
    \includegraphics[scale=0.7]{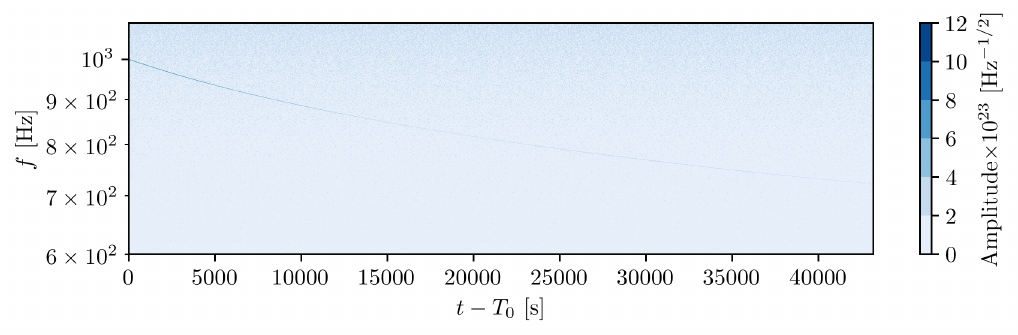} 
    \caption{\label{tcw_fig} Time-frequency evolution of an example signal in SFT data. The signal can be seen starting with $f_{\textrm{gw,0}}=1000$ Hz and getting dimmer as its frequency decreases with time.}
\end{figure}

The data used in this study are 8-second Han-windowed Short Fourier Transforms (SFTs), which are processed into either normalized power $\rho_j$ or number count $\nu_j$ as described by Oliver et al. \cite{oliver19}.
One example signal can be see in Figure \ref{tcw_fig}.

To search for signals in the data we use templates. These describe the corresponding time-frequency evolution track of potential signals in data. The power and number count statistics are weighed taking into account the antenna patterns and changing noise floors that modulate the signal amplitude. The statistics are then summed across the data bins selected by the track. If a signal is present, it will produce a strain in the data that causes an increase in power along is corresponding track. In order to quantify the significance of a given template, using either statistic $s$ we compute the critical ratio,
\begin{equation}
    \Psi_s=\frac{s-\langle s\rangle}{\sigma_{s}}\,,
\end{equation}
where $\langle s\rangle$ and $\sigma_{s}$ are the mean and variance of the corresponding statistic, computed as in \cite{oliver19}.

Two types of template banks have been employed in this study: (i) the cubic lattice template bank that arranges templates on a grid with equally spaced points in each dimension of the parameter space, and (ii) the uniform random template bank that samples templates from a uniform prior over the parameter space.
	
\section{Results \label{results}}

\begin{figure}[!h]
    \center
    \includegraphics[scale=1.1]{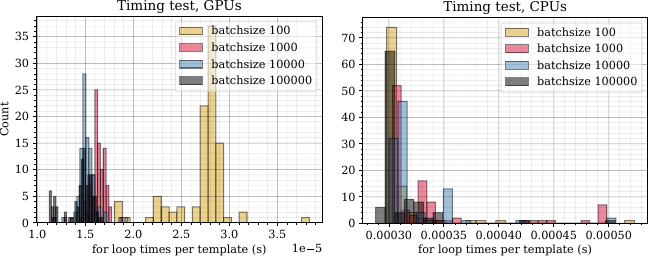} 
    \caption{\label{templates_timings_fig} Computing time per template obtained using either GPUs or CPUs with different batch sizes (templates computed simultaneously) tested on GPU NVIDIA Ampere A100 GPUs.}
\end{figure}

Thank to the usage of JAX, we have improved the computing time per template by a factor of 60 compared to the ATrHough method. Where the ATrHough method took around 1 ms per template, the new implementation on CPU takes only 0.3 ms. When using GPUs, this is reduced to only 15 µs per template. A 60-fold speedup such as this one is quite significant, since an analysis that would have needed 2 months of computing time would be reduced to just a single day. The timing results per template are shown on Figure \ref{templates_timings_fig}.

To evaluate the sensitivity obtainable with this method, we generated simulated signals consistent with a population of non-axisymmetric millisecond magnetar NSs uniformly distributed in all the sky, with random polarization, phase and orientation of the source NSs. 

The signals were created using the \texttt{lalpulsar\_Make-fakedata\_v5} code from LALsuite \cite{lalsuite}. A set of 100 signals per 19 distances between 0.1 and 3.1 Mpc in steps of 0.1 Mpc were generated. The $f_{\textrm{gw},0}$ of the injections was sampled from a uniform prior between 1000 Hz and 1200 Hz, with fixed $n=5$, $\tau=10000$ s and $\epsilon=10^{-2}$. The data duration was 24 h and the frequency range of the data from 300 to 1000 Hz. The signals were injected into colored Gaussian noise following the Advanced LIGO (Design) amplitude spectral density curve \cite{barsotti18}. 

Template banks with different numbers of templates, $\mathcal{N}$ = $\left[10^{4},10^{5},10^{6},10^{7}\right]$, were then generated to test the required density of templates for an effective search. The search coverage in frequency was the $(1000, 1200)$ Hz range, while for $n$ and $\tau$, the full theoretical ranges were used. Each template used the $T_0$, and sky position values of the injection being searched for.

Regarding the detection criteria, templates far away from the injected signal were excluded, and a threshold was set on $\Psi$ based on the distribution of critical ratios on only noise using the same amount of templates ($\Psi_{\textrm{noise max}}$). That is, an injection was considered detected if a template close to the injection had $\Psi>\Psi_{\textrm{noise max}}$.

\begin{figure}[!h]
    \center
    \includegraphics[scale=0.58]{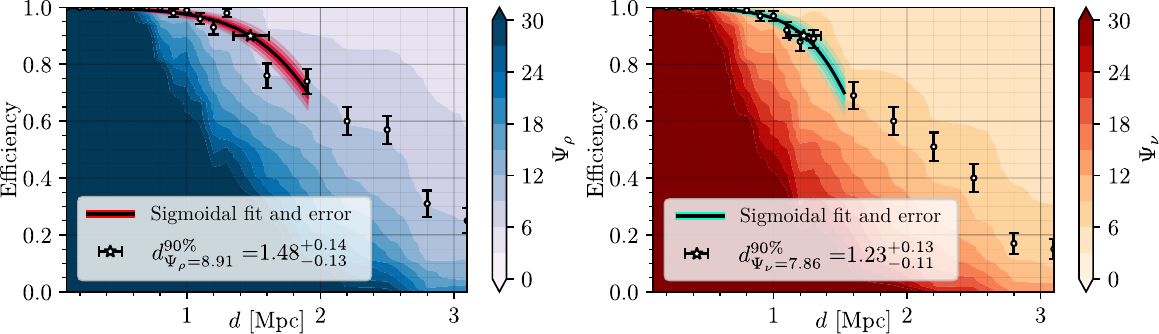} 
    \caption{\label{sigfits_fig} Efficiencies as a function of distance with their corresponding $d^{90\%}$ value for (left) normalized power and (right) number count using the uniform random template bank and $\mathcal{N}=10^7$. Each sigmoidal fit is shown with the corresponding 1, 2 and 3 sigmas error envelopes.}
\end{figure}

Efficiencies, $E$, were computed for different $\mathcal{N}$ and distance. Then, using a sigmoidal fit we estimated the 90\% detectability distances as shown in Figure \ref{sigfits_fig} for both the normalized power and number count statistics.

\begin{figure}[!h]
    \center
    \includegraphics[scale=0.58]{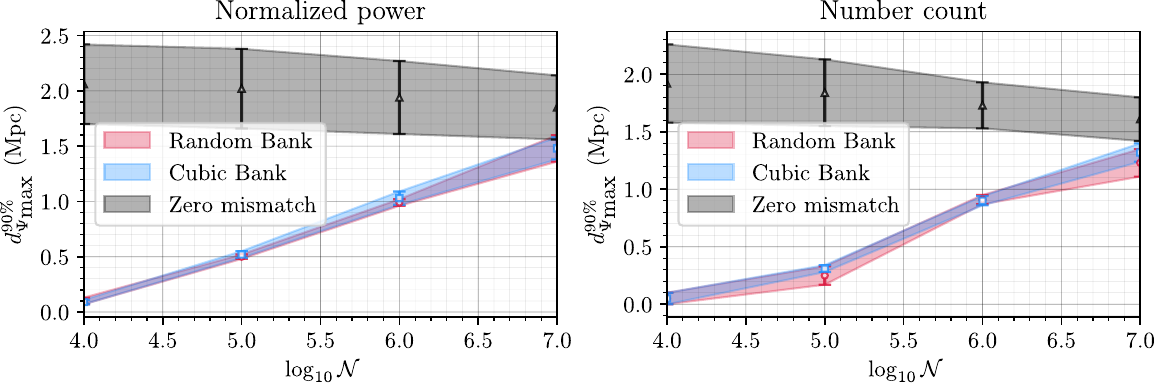} 
    \caption{\label{dist_results_fig} Detectability distances as a function of the number of templates. The red curves have been obtained using the uniform random template bank and the blue curves with the cubic template bank. The black curves are the detectability distances that are obtained with the tracks that follows exactly those of the injected signals, outlining the maximum obtainable sensitivity for this search.}
\end{figure}
The resulting $d^{90\%}$ values are presented in Figure~\ref{dist_results_fig}. As shown, the $d^{90\%}$ values continue to increase with the number of templates. At $\mathcal{N} = 10^7$, the distances approach the expected maximum sensitivity. Additionally, both template banks show nearly identical sensitivity. Furthermore, it is observed that distances derived from the number count statistic, $\nu$, are consistently 10\% to 15\% shorter than those calculated using the normalized power statistic, $\rho$. Finally, we can determine that a template bank density of $\mathcal{N} = 10^8$ to $10^9$ is sufficient to achieve maximum $d^{90\%}$ results in this scenario. Therefore, the number of templates $\mathcal{N}$ required is 1 to 2 orders of magnitude lower than what required by the ATrHough method.

\section{Conclusions \label{conclusions}}

We have implemented a new method for searches of long-transient gravitational waves from newborn remnants of binary neutron star mergers. Through studies of template bank configurations on campaign of simulated injected signals, we have determined the necessary template density to achieve nearly-optimal sensitivities in the studied scenario. The method  developed demonstrates sensitivities comparable to those of the previous ATrHough method, with the advantage of requiring fewer templates. Furthermore, the new codes run significantly faster per template compared to ATrHough with processing times up to 60 times faster thanks to the use of GPUs and JAX.
Our findings demonstrate the potential of GPU parallelization in significantly accelerating continuous wave searches. This work can be applicable to other long-duration GW signals \cite{tenorio2024jgc}.
%
%
\small  
%
\section*{Acknowledgments}   
%
JRM is supported by the Spanish  Ministerio de Ciencia, Innovación y Universidades (FPU22/01187). RT is supported by ERC Starting Grant No. 945155-GWmining, Cariplo Foundation Grant No. 2021-0555, MUR PRIN Grant No. 2022-Z9X4XS, MUR Grant “Progetto Dipartimenti di Eccellenza 2023-2027” (BiCoQ), and the ICSC National Research Centre funded by NextGenerationEU.. This work was supported by the Universitat de les Illes Balears (UIB); the Spanish Agencia Estatal de Investigación grants PID2022-138626NB-I00, RED2022-134204-E, RED2022-134411-T, funded by MICIU/AEI/10.13039/501100011033 and by the ESF+ and the ERDF/EU ; and the Comunitat Autònoma de les Illes Balears through the Servei de Recerca i Desenvolupament and the Conselleria d'Educació i Universitats with funds from the Tourist Stay Tax Law (PDR2020/11 - ITS2017-006), from the European Union - NextGenerationEU/PRTR-C17.I1 (SINCO2022/6719) and from the European Union - European Regional Development Fund (ERDF) (SINCO2022/18146). We gratefully acknowledge the computer resources at Artemisa and the technical support provided by the Instituto de Fisica Corpuscular, IFIC(CSIC-UV). Artemisa is co-funded by the European Union through the 2014-2020 ERDF Operative Programme of Comunitat Valenciana, project IDIFEDER/2018/048.
%

%

\begin{thebibliography}{}
\small

%
\bibitem{abbot16}{Abbott, B. P., et al. 2016, PRL, 116, 061102}
\bibitem{abbot17}{Abbott, B. P., et al. 2017, PRL, 119, 161101}
\bibitem{abbot18}{Abbott, B. P., et al. 2019, ApJ, 875, 160}
\bibitem{abbot21}{Abbott, B. P., et al. 2017, ApJ Lett., 851, L16}
\bibitem{abbot23}{Abbott, R., et al. 2023, PRX, 13, 041039}
\bibitem{alford12}{Alford, M. G., Schwenzer, K. 2014, ApJ, 781, 26}
\bibitem{barsotti18}{Barsotti, L., Fritschel, P., Evans, M., Gras, S. 2018, Tech. Rep. LIGO-T1800044-v3}
\bibitem{einstein16}{Einstein, A. 1916, Sitzungsber. Preuss. Akad. Wiss. Berlin (Math. Phys. ), 688}
\bibitem{jax2018github}{Bradbury, J., et al. 2018, http://github.com/google/jax}
\bibitem{lalsuite}{LIGO Scientific Collaboration, 2018, doi:10.7935/GT1W-FZ16}
\bibitem{lasky17}{Lasky, P., Sarin, N., Sammut, L. 2017, Tech. Rep. LIGO-T1700408}
\bibitem{lattimer2004}{Lattimer, J. M., Prakash, M. 2004, Science, 304, 536--542}
\bibitem{misner73}{Misner, C. W., Thorne, K. S., Wheeler, J. A. 1973, Gravitation, W. H. Freeman, San Francisco}
\bibitem{oliver19}{Oliver, M., Keitel, D., Sintes, A. M. 2019, PRD, 99, 104067}
\bibitem{reed21}{Reed, B. T., Deibel, A., Horowitz, C. J. 2021, ApJ, 921, 89}
\bibitem{tenorio2024jgc}{Tenorio, R., Mérou, J. R., Sintes, A. M. 2024, arXiv:2411.18370}
\bibitem{zimmerman79}{Zimmerman, M. and Szedenits, E., 1979, PRD 20, 351}
%
\end{thebibliography}
\end{document}